\documentclass[12pt,preprint]{aastex}
\usepackage{epsfig}
\usepackage{graphics}
\shortauthors{}
\shorttitle{X-ray emission from M31$^\ast$}

\begin{document}

\title{The Murmur of The Hidden Monster: {\sl Chandra}'s Decadal View of The Super-massive Black Hole in M31}
\author{Zhiyuan Li\altaffilmark{1}, Michael R.~Garcia\altaffilmark{1},
   William R.~Forman\altaffilmark{1}, Christine Jones\altaffilmark{1}, 
   Ralph P.~Kraft\altaffilmark{1}, Dharam V.~Lal\altaffilmark{1},
  Stephen S.~Murray\altaffilmark{1,2}, Q.~Daniel
  Wang\altaffilmark{3}
\altaffiltext{1}{Harvard-Smithsonian Center for Astrophysics, 60
  Garden Street, Cambridge, MA 02138; zyli@cfa.harvard.edu}
\altaffiltext{2}{Johns Hopkins University, 3400 North Charles St.,
   Baltimore, MD 21205}
\altaffiltext{3}{Department of Astronomy, University of Massachusetts,
  710 North Pleasant Street, Amherst, MA 01003}}

\begin{abstract}
The Andromeda galaxy (M31) hosts a central
super-massive black hole (SMBH), known as M31$^\ast$, which is remarkable for its mass
($\sim$$10^8{\rm~M_\odot}$) and 
extreme radiative quiescence. Over the past decade, the {\it Chandra X-ray
  observatory} has pointed to the center of M31 $\sim$100 times
and accumulated a total exposure of $\sim$900 ks. Based on these
observations, we present an X-ray study of a highly variable source
that we associate with M31$^\ast$ based on positional coincidence.
We find that M31$^\ast$
remained in a quiescent state from late 1999 to 2005, exhibiting
an average 0.5-8 keV luminosity
$\lesssim$$10^{36}{\rm~ergs~s^{-1}}$, or only $\sim$$10^{-10}$ of its
Eddington luminosity.
We report the discovery of an outburst that occurred
on January 6, 2006, during which M31$^\ast$
radiated at $\sim$$4.3\times10^{37}{\rm~ergs~s^{-1}}$. After the outburst, M31$^\ast$ entered
a more active state that apparently lasts to the present, which is characterized by frequent flux variability
around an average 
luminosity of $\sim$$4.8\times10^{36}{\rm~ergs~s^{-1}}$. 
These flux variations are similar
to the X-ray flares found in the SMBH of our Galaxy (Sgr A$^\ast$), 
making M31$^\ast$ the second SMBH known to exhibit recurrent
flares. 
Future coordinated X-ray/radio
observations will provide useful constraints on the physical origin of
the flaring emission and help rule out a possible stellar origin of the X-ray
source.
\end{abstract}

\section{Introduction} {\label{sec:intro}}
Most, if not all, galaxies with a stellar bulge are thought to harbor a super-massive black
hole (SMBH) in their nuclei. Accretion onto and feedback (i.e., both
radiative and mechanical energy output) from the SMBH is one
of the fundamental astrophysical processes that govern galaxy
evolution. 
Compared to their
high-redshift counterparts, most SMBHs in the local universe, when observed,
are found to be radiatively quiescent (e.g., Zhang et al.~2009; Gallo
et al.~2010),
and are often dubbed low-luminosity active galactic nuclei (LLAGNs;
cf.~Ho 2008). By analogy to Galactic black
hole binaries in a low/hard state (cf. McClintock \& Remillard 2006), 
LLAGNs are generally thought to be powered by 
radiatively inefficient, advection-dominated accretion and/or outflow 
(Narayan \& Yi 1994; Blandford \& Begelman; Quataert \& Gruzinov 2000)
that operate  at very sub-Eddington accretion
rates\footnote{less than a few percent of the Eddington mass accretion rate $\dot{M}_{\rm Edd}{\equiv}10L_{\rm
Edd}/c^2$, where $L_{\rm Edd}=1.3\times10^{38}(M_{\rm BH}/M_\odot){\rm~erg~s^{-1}}$  is the Eddington luminosity}. 
Albeit often subject to instrumental limitations as a consequence of their
radiative quiescence, studies of LLAGNs 
have important implications for accretion physics, fueling and
feedback mechanisms, and black hole growth over cosmic time. 

A well-known LLAGN is the SMBH in our Galaxy,
Sgr A$^\ast$ (cf. Melia \& Falcke 2001), which has an extremely quiescent bolometric luminosity of
$\sim$$3\times10^{-9} L_{\rm Edd}$ (for its mass of
$\sim$$4\times10^6{\rm~M_\odot}$; Ghez et al.~2003), a factor of $10^2-10^6$ lower
than the inferred values for most LLAGNs (e.g., Zhang
et al.~2009; Ho 2009). More unusual about Sgr A$^\ast$ is its
flaring emission detected in X-ray, infrared and radio bands (Baganoff et al.~2001; Genzel
et al.~2003; Zhao et al.~2003). 
In particular, the X-ray flares show the greatest
variability, with hour-timescales and peak fluxes reaching $\sim$10-100 times the
quiescent level (e.g., Baganoff et al.~2001; Porquet et al.~2003, 2008). 
In contrast, most LLAGNs show only a modest level of X-ray
variability (e.g., Ptak et al.~1998).

Until now, Sgr~A$^\ast$ has remained the only LLAGN found to
exhibit recurrent flares. 
In this {\sl Letter}, we report the discovery of flaring X-ray emission from
the SMBH in M31, based on the
most up-to-date data obtained from the {\sl
  Chandra X-ray Observatory} (Weisskopf et al.~2002).
A distance of 780 kpc is adopted for M31.
We quote 1$\sigma$ uncertainties throughout this work.

\section{The SMBH in M31} {\label{sec:M31}}
We begin with a brief overview of the most relevant observational aspects
of the SMBH in M31.
The center of M31 exhibits the well known double-peaked
feature in the optical and near-UV (the so-called double nuclei, P1
and P2; Lauer et al.~1993),
which is interpreted as an eccentric disk of primarily
old stars orbiting around the SMBH (Tremaine 1995). 
The latter is embedded in P2\footnote{P2 is referred to as the peak
  that is fainter
in the optical but brighter in the near-UV. The UV-peak is further designated as P3 according
to Bender et al.~(2005)}, with an inferred dynamical mass of $1.4^{+0.9}_{-0.3}\times10^8 {\rm~M_{\odot}}$ (Bender et al.~2005).

Possible counterpart of the SMBH has been detected only in the radio and X-ray bands to date. 
Using Very Large Array (VLA) observations at 8.4 GHz, 
Crane, Dickel \& Cowan (1992) found an unresolved source, named
M31$^{\ast}$ following the convention of Sgr A$^\ast$, at a position
coincident with P2. A marginal 8.4 GHz flux
variation was reported by Crane et al.~(1993).
Based on a {\sl Chandra}/HRC observation, Garcia et al.~(2005)
claimed a 2.5 $\sigma$ detection of X-ray emission from the position
of P2. 
Based on a set of {\sl Chandra}/ACIS observations taken before 2006, Li, Wang \& Wakker (2009)
determined an intrinsic luminosity of $\sim1.2\times10^{36}
{\rm~ergs~s^{-1}}$ at the position of P2, which places
a firm upper limit to the (quiescent) X-ray emission from the
SMBH, corresponding to only $\sim$$10^{-10} L_{\rm Edd}$. 
With more recent {\sl Chandra}/HRC observations, Garcia et al.~(2010)
reported X-ray flux variations of P2 by 
a factor of $\sim$3 on a timescale of days and a factor
of $\gtrsim$10 in a year. They also detected M31$^\ast$ at 5 GHz with VLA, 
finding no significant flux variation among four observations
near the end of 2004. 

\section{Data preparation} {\label{sec:data}}
In this work, we utilized 98 {\sl Chandra} observations taken between 
October 13, 1999
and March 5, 2010, which irregularly sampled the time domain. 
Typically, in each year
there were $\sim$10 observations, with two successive
observations separated by hours to months.
The full dataset consists of 58 ACIS observations and 40 HRC observations.
The HRC best employs the
angular resolution of the {\sl Chandra} High Resolution Mirror Assembly (HRMA),
but essentially lacks spectral resolution. The ACIS has  
higher broad-band effective area than the HRC, but
slightly undersamples the point-spread function (PSF) of the HRMA.
The majority of the ACIS observations have exposures of $\sim$5 ks
and the aimpoint on the I3 CCD. 
The HRC observations, all taken with the I-array, have more diverse
exposures ranging from 1 to 50 ks.  
In all cases, the nucleus of M31 was placed within $1^\prime$ of
the aimpoint, and hence was observed with optimal angular resolution.

We reprocessed the data using CIAO version 4.2 and the corresponding
calibration files. We followed the standard procedure of data reduction, except that we
turned off the sub-pixel position randomization for the ACIS events. 
To maximize
the temporal coverage, we performed no filtering on time intervals of high particle background.  
We verified that,
in all observations, the instrumental background
is negligible compared to the emission from
the central few arcseconds of M31.
The total ACIS and HRC exposures are 305 ks and 571 ks, respectively.
We considered only ACIS events in the 0.5-8 keV range and HRC events
in the {\sl PI}
range of 10-350, effectively excluding the majority of instrumental events.

A crucial step for our study is to calibrate the relative astrometry
among individual observations, which was done by matching the centroids
of about 30 discrete, relatively bright sources located within 1$^\prime$ (but outside
5$^{\prime\prime}$) of the nucleus. The resulting accuracy is
typically better than 0\farcs1. For each observation, we
also generated an exposure map, primarily to account for the gradual change in
the detector effective area.
 An absorbed power-law spectrum,
with a photon-index of 1.7 and an absorption column density
$N_{\rm H}=10^{21}{\rm~cm^{-2}}$, 
was adopted to calculate spectral weights when
producing the exposure maps. The energy-dependent difference of
the effective area between
the ACIS-I, ACIS-S3 and HRC-I, 
was taken into account, assuming the above incident spectrum, so that
the quoted count rates throughout this work refer to ACIS-I.
 

\section{Analysis and results} {\label{sec:results}}
\subsection{X-ray sources in the nuclear region of M31}  
\begin{figure}
\centerline{
\epsfig{figure=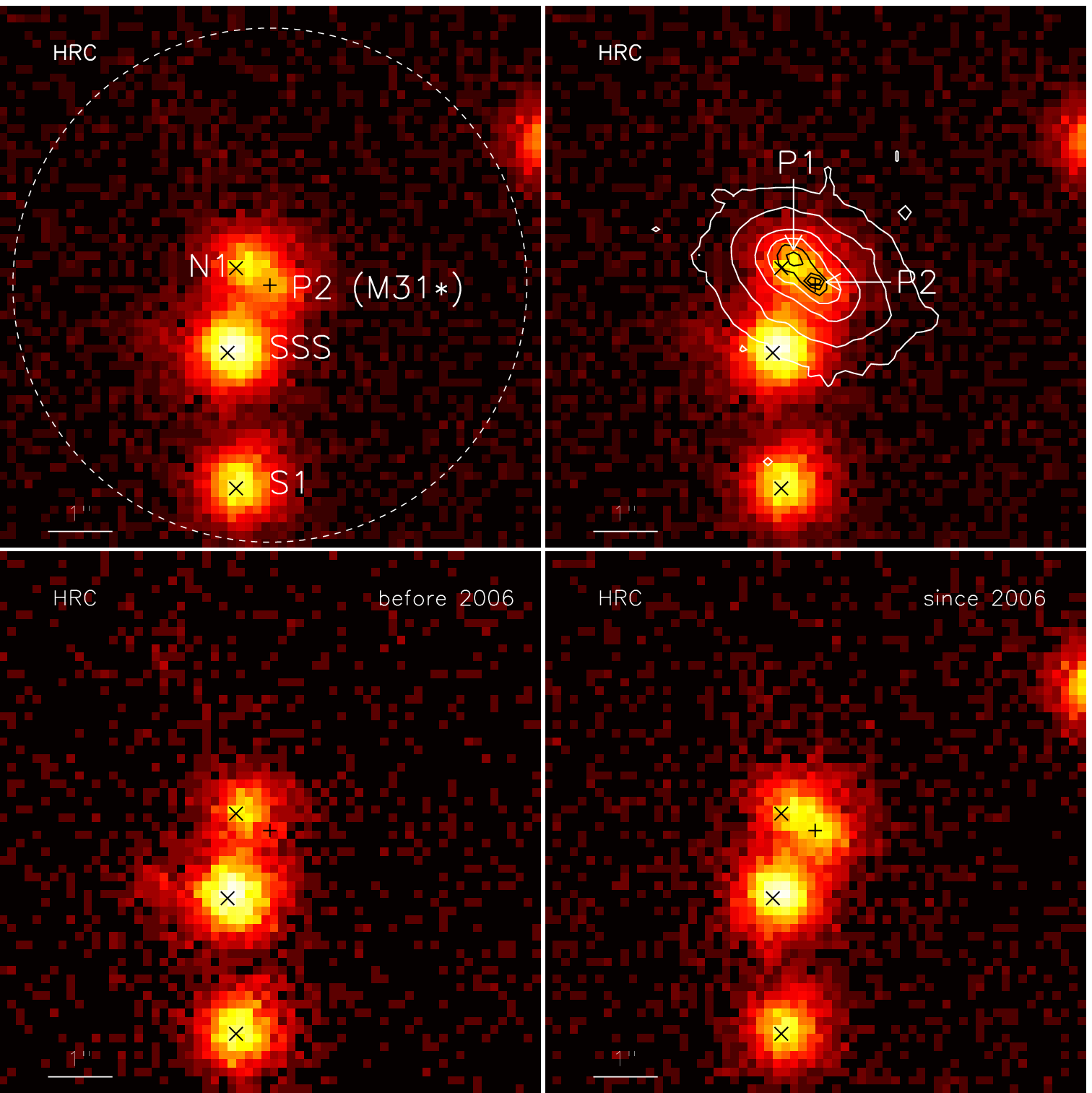,height=0.7\textheight,angle=0,clip=}
}
\caption{{\sl Chandra}/HRC count images of the
  $8^{\prime\prime}{\times}8^{\prime\prime}$ nuclear region of M31, 
  from stacking all observations ({\sl top left} and {\sl top right}), 
  observations taken before 2006 ({\sl bottom
  left}) and observations taken since 2006 ({\sl bottom right}), respectively. The HRC pixel size is 0\farcs1318.
  The position of M31$^\ast$ is
  marked by a `+' and labelled as P2. The three extranuclear
  sources, labelled as, from north to south, N1, SSS and S1,
  are marked by crosses. The dashed circle encloses the region for fitting. 
{\sl HST}/ACS F330W intensity contours are plotted
  in the {\sl top right} panel, highlighting the double nuclei (P1 and
  P2) in the optical. Note that P2 becomes significantly brighter since 2006.}
\label{fig:HRC_nuc}
\end{figure}

{\label{subsec:image}}
It is known that several bright X-ray sources are present in the
nuclear region of M31, challenging a clean isolation of the emission
from M31$^\ast$ (Garcia et al.~2000; Li et al.~2009).
We first examine the HRC observations, since they have the better angular resolution
(PSF FWHM $\approx$ 0\farcs4). 
Fig.~\ref{fig:HRC_nuc}a and Fig.~\ref{fig:HRC_nuc}b show an image of 
the nuclear region, stacking all HRC observations on the original pixel scale of 0\farcs1318.
Four X-ray sources are immediately identified within 4$^{\prime\prime}$
of the nucleus. 
Following Garcia et al.~(2005) and Li et al.~(2009), we label these
sources in Fig.~\ref{fig:HRC_nuc}a as:
P2 for the source positionally coincident with the nucleus, N1 for the source just $\sim$0\farcs5 northeast of P2, 
SSS for the source $\sim$1\farcs5 south of N1, and S1 for the source
$\sim$2$^{\prime\prime}$ farther south.
The positional coincidence between P2 and the nucleus is demonstrated 
in Fig.~\ref{fig:HRC_nuc}b, in which {\sl HST}/ACS intensity contours of the
optical double-nuclei of M31 (Lauer et al.~1993) are shown.  
We have registered the HRC and {\sl HST}/ACS images by matching 
common extra-nuclear sources, resulting in an uncertainty of
$\sim$0\farcs2 (Garcia et al.~2010).
The other three sources are most likely stellar objects (Kong et
al.~2002; Di Stefano et
al.~2004; Li et al.~2009).

Visual inspection of individual HRC observations reveals that all four
sources show some degree of flux variation, with P2 varying most significantly. 
This can be illustrated by
dividing the HRC data into two subsets, the first half consisting of
observations taken before 2006 (Fig.~\ref{fig:HRC_nuc}c; a total
exposure of 256 ks) and the
second half consisting of observations taken since 2006
(Fig.~\ref{fig:HRC_nuc}d; a total exposure of 315 ks).
Using the three extra-nuclear sources as a reference, it is obvious
that P2 has become 
significantly brighter since 2006. 
We note that two more sources can be identified in
Fig.~\ref{fig:HRC_nuc}. The fainter one only appears before 2006, located
immediately east of SSS; the brighter one only appears since 2006,
located at the upper right (northwest) corner 
of the images. Neither of these two
transient sources affects the following analysis and hence will not be
discussed further.

\begin{figure}
\centerline{
\epsfig{figure=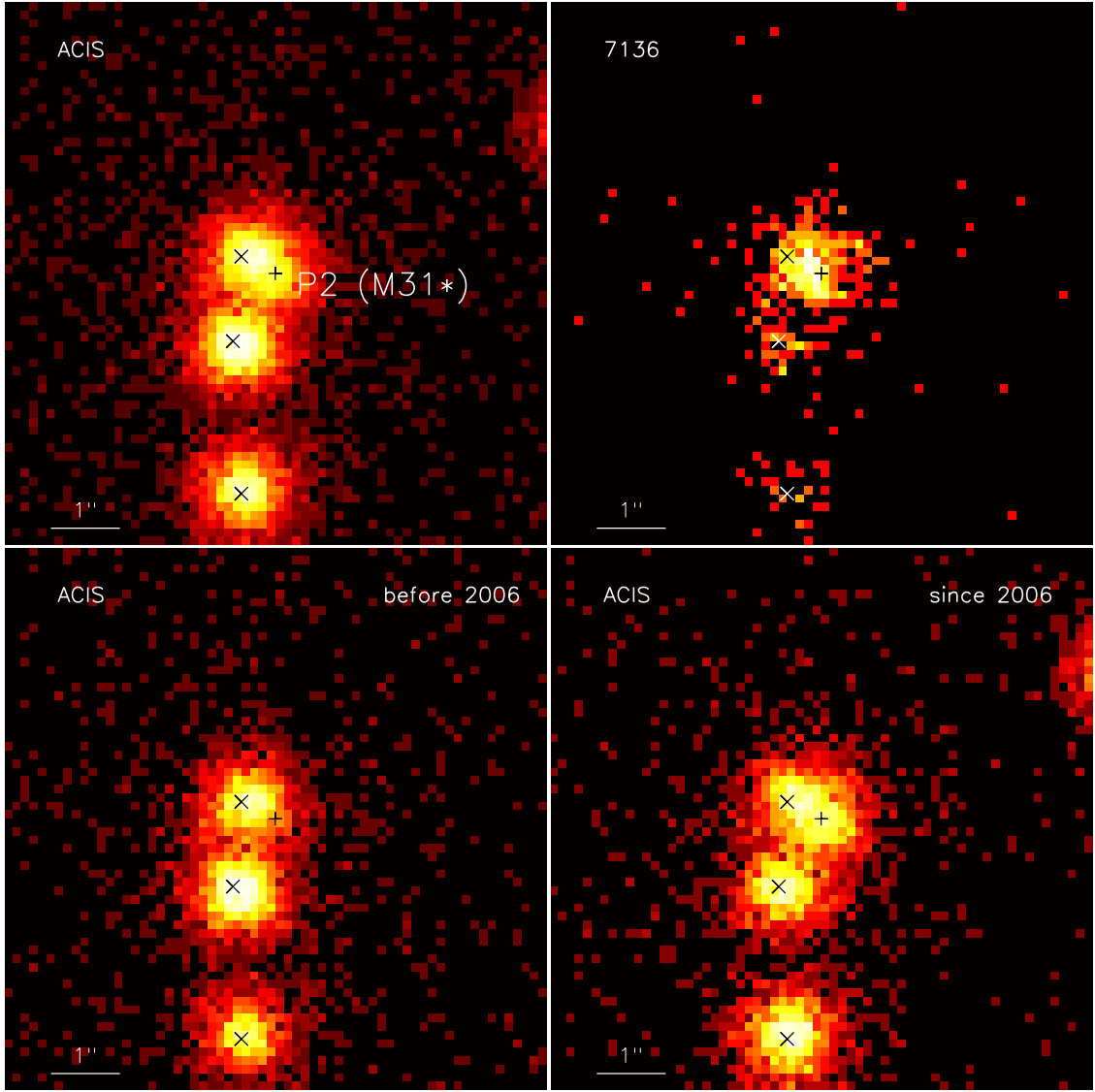,height=0.7\textheight,angle=0,clip=}
}
\caption{{\sl Chandra}/ACIS 0.5-8 keV count images of the
  $8^{\prime\prime}{\times}8^{\prime\prime}$ nuclear region of M31,
  from stacking all observations ({\sl top left}), 
  ObsID 7136 (taken on January 2, 2006) alone ({\sl top right}), stacking observations taken before 2006 ({\sl bottom
  left}) and stacking observations taken since 2006 except for ObsID 7136 ({\sl
  bottom right}), respectively. The pixel size is 0\farcs123, i.e.,
  1/4 of the ACIS pixel size.
 The sources are marked in the same way as in
  Fig.~\ref{fig:HRC_nuc}. Note the outburst of P2 captured by ObsID 7136.}
\label{fig:ACIS_nuc}
\end{figure}

We now turn to the ACIS observations. An image stacking all 58 ACIS observations 
is shown in Fig.~\ref{fig:ACIS_nuc}a, on a pixel scale
of 0\farcs123 (i.e., one fourth of the original ACIS pixel size) to take
advantage of the sub-pixel positioning information that results from telescope dithering
and to approximately match the HRC pixel scale. 
With a PSF FWHM of $\sim$0\farcs6, the four sources are again 
clearly identified. 
Visual inspection of individual ACIS observations consistently
reveals the flux variability in P2.
Most dramatically, in an observation taken on January 6, 2006 (ObsID
7136, with an exposure of 5 ks), P2 appears to be by far the brightest
source in the field (Fig.~\ref{fig:ACIS_nuc}b), indicating that an
outburst has been fortuitously captured.
We similarly present two subsets of the
ACIS data, i.e, observations taken before 2006
(Fig.~\ref{fig:ACIS_nuc}c; a total exposure of 149 ks) and since 2006 (but excluding ObsID
7136; Fig.~\ref{fig:ACIS_nuc}c; a total exposure of 152 ks).
Again, the two subsets show a marked difference in the brightness
contrast between P2 and the three extra-nuclear sources.

\subsection{A decadal light curve} {\label{subsec:lc}}
To obtain a more quantitative view of the flux variability, we
construct a light curve for P2.
Due to the proximity of the four sources, in particular between N1 and
P2, it is crucial to account for the mutual PSF scattering.
We adopt a two-dimensional fitting procedure, similar to that employed by Li et al.~(2009), 
to simultaneously determine the fluxes for all four sources in each observation. 
Specifically, we take the following steps: 

(i) From the circumnuclear sources used for calibrating the astrometry
(\S~\ref{sec:data}), we further select relatively isolated ones
($\sim$10), i.e., those without neighboring sources within 4$^{\prime\prime}$.
For each observation, we obtain a local PSF image by stacking the selected sources after
centroiding. The average PSF for the summed image
(be it over a given detector, a given energy range or a given epoch) is then obtained by stacking individual PSF
images of the constituent observations.

(ii) With the CIAO tool {\sl Sherpa}, we fit the summed image, using a 
model consisting of four point-like sources and a constant local
background. Each source is represented by a delta
function, whose center and normalization are free parameters. This model
is convolved with the average PSF. Only the central 4$^{\prime\prime}$ from the nucleus is
considered in the fit (outlined by the circle in
Fig.~\ref{fig:HRC_nuc}a). Given the limited counts in each 
unbinned pixel, the best-fit is obtained by minimizing the C-statistic (Cash 1979).
The fit is robust in determining the source centroids,
the chief goal of this step.

(iii) For each observation, we repeat the above fitting procedure,
fixing the centroids of all four sources as determined from the summed image. 
The fit then allows us to determine the flux and its
uncertainty, corrected for the
local effective exposure, for each source in each observation.
Such a procedure not only maximizes the counting statistics, but also properly 
accounts for the propagation of the dominating error term arising from
the mutual PSF scattering.

\begin{figure}
\centerline{
\epsfig{figure=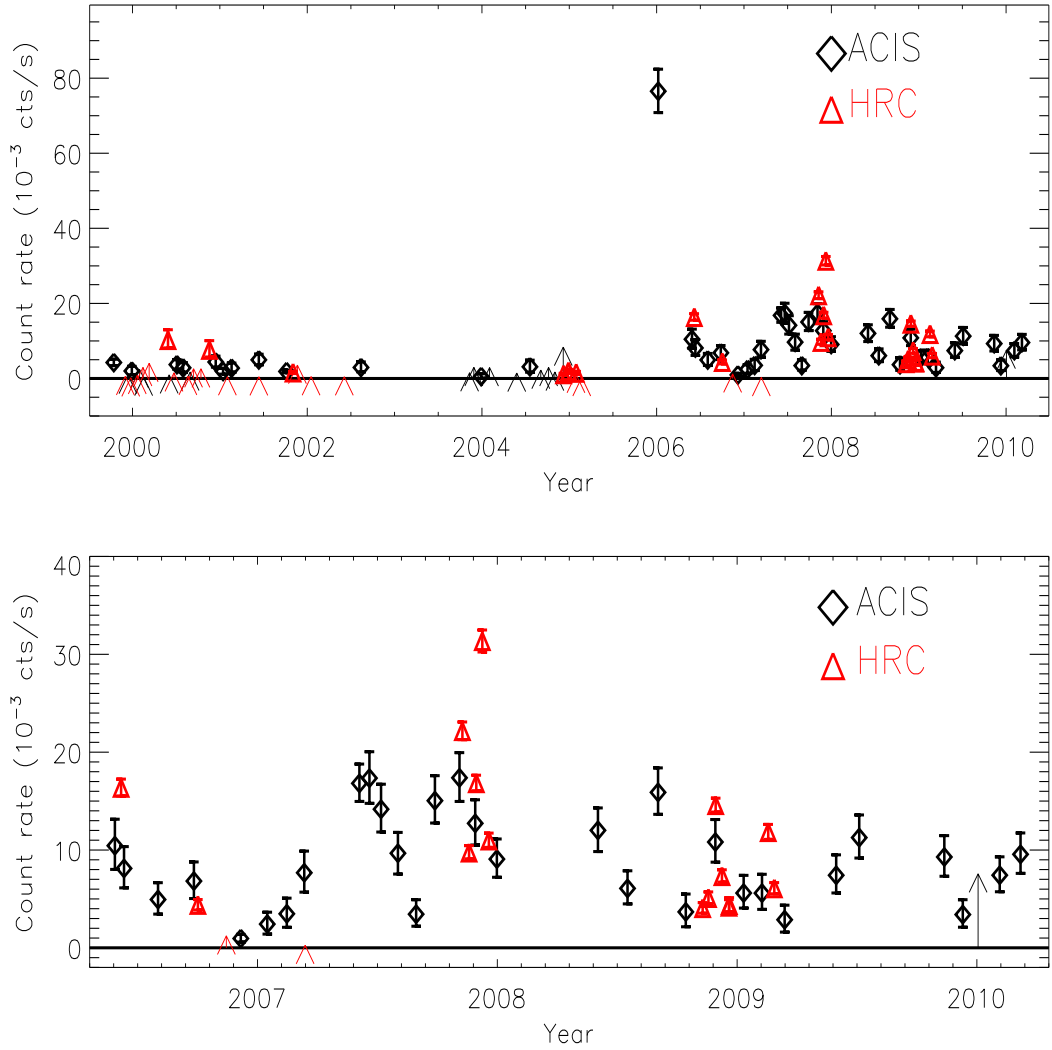,height=0.7\textheight,angle=0,clip=}
}
\caption{{\sl Upper panel}: The decadal X-ray light curve of
M31$^\ast$. 
An outburst is captured on January 6, 2006, followed by an epoch of
frequent flux variation, shown in the 
{\sl Lower panel}. Arrows represent measurements whose 1$\sigma$ lower limit is consistent with zero flux. See text for details.
}
\label{fig:lc}
\end{figure}

The resultant decadal light curve of P2 (Fig.~\ref{fig:lc}) shows two prominent features.
First,  an 
outburst, preceded by a 6-yr epoch of quiescence, appears on January 6, 2006, during which the flux of P2
reaches a value of
$(76.5\pm5.9)\times10^{-3}{\rm~cts~s^{-1}}$. 
Second, after the outburst, P2 apparently enters a more active state,
characterized by frequent flux variations.
Summing the observations taken before 2006 and repeating the above
fitting procedure, we obtain an average flux of $(0.3\pm0.1)\times10^{-3}{\rm~cts~s^{-1}}$.
Similarly, for the observations since 2006 (but
excluding ObsID 7136), we find an average flux of
$(6.9\pm0.3)\times10^{-3}{\rm~cts~s^{-1}}$.
Therefore, the observed flux of the outburst is
$\sim$250 times the quiescent level before it,
whereas the long-term average flux increases by a factor of $\sim$23 
after the outburst. For comparison, the three extra-nuclear
sources\footnote{We will present elsewhere a detailed timing analysis
  for these three sources as well as other stellar X-ray sources in
  the M31 bulge.}
show a ratio of fluxes since and before 2006 of $\sim$1.1 (N1),
$\sim$0.7 (SSS), $\sim$1.5 (S1), respectively, all indicating a nearly
constant long-term flux; their maximum-to-mean flux ratios are $\sim$3-5,
also suggesting that the flux variation seen in P2 is exceptional.

It is impractical to perform a spectral analysis for P2,
due to the limited number of counts in most observations and the
contamination of scattered photons from the other
sources whose spectral information is also uncertain. 
Instead, by dividing the ACIS counts into a soft band and a hard band, 
we estimate the hardness ratio,
defined as $HR=(I_{\rm 0.5-2~keV}-I_{\rm 2-8~keV})/(I_{\rm 0.5-2~keV}+I_{\rm 2-8~keV})$, to be  
$-0.11\pm1.30$, $-0.76\pm0.04$ and $-0.50\pm0.04$ before, during and
after the outburst, respectively.
While in the first epoch $HR$ is essentially indeterminate, 
it appears that the emission during the outburst is softer than that in the last epoch.
For comparison, a Galactic foreground-absorbed ($N_{\rm H}=7\times10^{20}{\rm~cm^{-2}}$) power-law spectrum with a
photon-index of 1.8 (2.6) would result in $HR=-0.50 (-0.76)$.
We thus convert the observed count rates into 0.5-8 keV intrinsic luminosities 
of 0.2, 43 and 4.8$\times10^{36}{\rm~ergs~s^{-1}}$ for the three
epochs, by assuming a power-law spectrum with a photo-index of 1.8,
2.6 and 1.8, respectively.

We find no statistically significant intra-observation flux variation
for the outburst. We note that a typical 5-ks long observation may probe flux
variations arising from a region of $\sim$$5\times10^{-5}$ pc in diameter, roughly
4 times the Schwarzchild radius of the SMBH. 
Unfortunately, ObsID 7136 is the most isolated
observation: no other observation was taken
before or after it within several months. Therefore we cannot 
rule out that P2 stayed at high flux levels for up to months
between 2005-2006. The fact that we have found only one outburst in
98 observations indicates a $\sim$1\% duty cycle for outbursts of this kind.


\section{Discussion} {\label{sec:disc}}
\subsection{P2 as M31$^\ast$}
As in previous studies (Garcia et al.~2005, 2010; Li et al.~2009), our
identification of P2 as the X-ray counterpart of M31$^\ast$ rests on
positional coincidence. The uncertainty of the registration among the X-ray,
optical and radio images is $\sim$0\farcs1-0\farcs2, corresponding to
a linear size of $\lesssim$1 pc, and the chance coincidence of
an interloping X-ray source is only $\lesssim$1\% (Garcia et al.~2010). 
On the other hand,
within 1 pc of the SMBH, members of the eccentric stellar disk (Lauer et al.~1993; Tremaine
1995) almost certainly produce some 
X-ray emission, which is however difficult to reliably
quantify due to the underlying extremely high stellar density. 
Indeed, the absolute luminosities of the outburst and
the subsequent individual detections are only modest
($\lesssim$a few $10^{37}{\rm~erg~s^{-1}}$),
and by face value, can be
attributed to an X-ray binary, e.g., containing an accreting
stellar-mass black hole. Most known Galactic black hole binaries show outbursts 
at a substantial fraction of their $L_{\rm Edd}$, and in rare cases, some of them also show
complex flux variability  
(e.g., McClintock \& Remillard 2006). 
Hence before addressing the nature of the outburst in the context of
M31$^\ast$, we note that a stellar origin of 
P2 cannot be completely ruled out.

\subsection{Origin of the outburst}

The detected outburst adds M31$^\ast$ to a short list of non-active SMBHs
that have shown strong X-ray flux variations. Early {\sl ROSAT} observations
revealed a handful of X-ray outbursts associated with the
nuclei of optically non-active galaxies (cf. Komossa 2002), which are
interpreted in terms of tidal disruption of a star passing by the SMBH
(Rees 1988). The absolute luminosities of these outbursts are
$\sim$$10^{42}-10^{44}{\rm~erg~s^{-1}}$, a substantial fraction of
$L_{\rm Edd}$
of the associated SMBH. Moreover, these outbursts are transient rather
than recurrent. These are to be contrasted with the M31$^\ast$ outburst, whose
luminosity is $\sim$10$^{-8.5}$ $L_{\rm Edd}$ and which is followed by recurrent flux variations, albeit
with smaller amplitudes. Hence the M31$^\ast$ outburst
is unlikely to be triggered by tidal
disruption event.

The outburst is more reminiscent of the strongest X-ray flares seen in
Sgr A$^\ast$ (Porquet et al.~2003, 2008). 
The M31$^\ast$ outburst and Sgr A$^\ast$ flares are similar in strength ($\gtrsim$100 times relative to
the quiescent level), spectral softness and low incidence rate.  
Hence it is reasonable to speculate that the outburst shares the same physical
origin with the Sgr A$^\ast$ flares. 
In this regard, the individual detections since
the outburst are likely flares of smaller amplitudes, again similar
to the smaller, but more frequent, flares seen in Sgr A$^\ast$.
However, it appears that 
the outburst either has triggered or signaled the smaller flares, and the
post-outburst mean flux level became more than 10 times the mean flux before the outburst.
No such trend has been reported in the long-term X-ray emission from Sgr A$^\ast$. 

At present, the physical origin of the Sgr A$^\ast$ flares remains elusive and most
proposed models are phenomenological (e.g., Yuan, Quataert \& Narayan 2004;
Liu, Melia \& Petrosian 2006; Maitra, Markoff \& Falcke 2009).
In a magnetohydrodynamical model analogous to that developed for 
the phenomena of coronal mass ejection in our Sun (Lin \& Fobes 2000),  
Yuan et al.~(2009a) interpreted the flares as episodic ejection of
relativistic plasma blobs (``episodic jets'') inflated by magnetic 
field reconnection in the inner region of the accretion flow. Relativistic particles
are accelerated in the reconnection current sheet
and are responsible for the X-ray flares. After Sgr A$^\ast$, M31$^\ast$ is only the
second SMBH found to show flaring emission. A comparative study of 
the two SMBHs should yield important constraints to the flare modeling.
Interestingly, M31$^\ast$ and Sgr A$^\ast$ are also the two least
active SMBHs known. This refreshes the issue of whether flares are
fundamentally related to the lowest accretion rates of LLAGNs, as
discussed by Yuan et al.~(2004).


Regardless of the exact physical process that produces the active X-ray
emission from M31$^\ast$, it is expected that the average radio
emission from M31$^\ast$ has also substantially increased since 2006, 
by analogy to the associated X-ray/radio flares in Sgr A$^\ast$ (e.g.,
Marrone et al.~2008). This is also suggested by the so-called 
fundamental plane of black hole activity (Merloni, Heinz \& Di Matteo 2003;
see also Yuan et al.~2009b), which empirically relates the X-ray
luminosity, radio luminosity and mass of an accreting black hole. 
We investigated archival VLA observations taken since 2006, but found
 no useful constraints on the radio variability of M31$^\ast$,
due to poor data quality.
New coordinated EVLA observations with much enhanced sensitivities will
shed light on the origin of the X-ray flares in M31$^\ast$. 
Simultaneous X-ray/radio detection of variability in P2 will
help rule out an X-ray binary, whose radio flux is expected to be well
 below the sensitivity limit of EVLA.



\vspace{0.5cm}
We are grateful to Feng Yuan for valuable discussions.   
ZL thank Jun Lin and Jeffrey McClintock for useful comments, and the {\sl Chandra} calibration team, in particular, Jennifer
Posson-Brown, Frank Primini and  Aneta Siemiginowska, for
suggestions on the usage of HRC data and {\sl Sherpa}. 
This work is supported by SAO grants GO9-0100X and GO0-11098.

\clearpage
\clearpage

\end{document}